\documentclass[%
reprint,
 amsmath,amssymb,
 aps,
]{revtex4-2}

\usepackage{graphicx}
\usepackage{dcolumn}
\usepackage{bm}
\usepackage{amsmath}
\usepackage{nccmath}
\usepackage[dvipdfmx]{hyperref}
\usepackage[dvipdfmx]{xcolor}
\usepackage{mathtools}
\usepackage{natbib}

\begin{document}

\preprint{APS/123-QED}

\title{Dynamical mechanism of fusion hindrance in heavy ion collisions}

\author{Shota Amano$^{1}$, Yoshihiro Aritomo$^{1}$ and Masahisa Ohta$^{2}$\\
\scriptsize{$^{1}$Kindai University Higashi-Osaka, Osaka 577-8502, Japan}\\
\scriptsize{$^{2}$Konan University Kobe, Hyogo 658-8501, Japan}\\
\scriptsize{e-mail: 2144340401y@kindai.ac.jp}\\
}

\date{\today}

\begin{abstract}
\noindent{\bf Background:} In the fusion process, the investigation of the reaction dynamics in the time evolution of the nuclear configuration is necessary. The neck parameter $\epsilon$ which is one of the parameters representing the nuclear configuration in the two center shell model is important in fusion owing to the nucleons transferring through the neck. The time evolution of the neck has not been discussed in detail, but is crucial for fusion cross section in the assessment of new elements synthesis.
\\
{\bf Purpose:} The dynamical analysis for the fusion hindrance under the neck formation on the nuclear deformation space has been done. The fusion probability $P_\text{CN}$ considering the different denecking motion and the fusion hindrance are discussed.\\
{\bf Method:} The calculations were performed using the dynamical model of nucleus-nucleus collisions based on the multidimensional Langevin equations.\\
{\bf Results:} The formation of the neck bridge at the approaching stage is found to be crucial to the fusion hindrance.
It is clarified that
the inner barrier appears owing to the change in the degree of mass asymmetry $\alpha$ with the relaxation of $\epsilon$.\\
{\bf Conclusions:} The fusion hindrance occurs because the inner barrier is formed by the early neck formation. The role of the neck parameter $\epsilon$ is critically important for the fusion dynamics.
\end{abstract}

\maketitle


\section{Introduction}

The heavy ion reaction has been categorized into several processes. In the first step, the projectile and target nucleus stick each other after overcoming the interaction barrier mainly from the Coulomb potential (capture process). Next, the system moves along the path toward forming the compound nucleus (CN) (fusion process) and the dominant part of the event evolves to the reseparation after exchanging some amount of nucleons (quasifission (QF) or deep inelastic collision (DIC) processes).

The capture cross section is defined as a sum of the QF, DIC and CN cross section, and is measurable quantity by experiments. Theoretically, the capture-cross section can be defined by the total sum of the transport coefficient of the interaction barrier weighting the angular momentum factor $(2\it l+\text{1})$ corresponding to the impact parameter. The important thing is the identification of the fusion phenomena competing with QF and DIC. The estimation of the CN cross section is crucial for the prediction of the synthesis of superheavy elements. Because the superheavy element is identified as the evaporation residue (ER) of CN surviving from the dominant process of fission.

Many attempts to separate the capture cross section from QF and DIC, and to identify the CN cross section have been reported. In the early macroscopic dynamical model by Swiatecki \cite{WJSwiatecki1981}, CN cross section was estimated taking into account of the neck degree of freedom. The fusion probability was also calculated by Smolkowski diffusion model \cite{PhysRevC.55.R1011} and the probability passing through the CN region of deformation space was estimated by the multidimensional Langevin equation \cite{doi:10.1063/1.55148}. Recently, the Langevin equation is used widely for the analysis of fusion and fission phenomena in superheavy mass region \cite{PhysRevC.99.051602,naderi2018influence,litnevsky2020formation,ZAGREBAEV2015257,ishizuka2020effect,miyamoto2019origin}.
Another approach to the CN cross section is based on the dinuclear system (DNS) model \cite{ADAMIAN199929,ADAMIAN2000233}, in which they pointed the importance of the neck behavior of the colliding system, and presented that the theoretical overestimation of fusion probability in heavier collision systems is refined by the proper treatment of the mass parameters for the neck degree of freedom.
%

It is necessary to investigate in detail the dynamics of the time evolution of the shape of nuclei.
Especially, the time evolution of the neck is one of the important factor \cite{ADAMIAN199929,PhysRevC.83.054620,PhysRevC.106.024610}, but the detail has not been discussed. Therefore, it is essential to analyze the evolution of the neck parameter, its role, and the contribution in the fusion process.
%

In the present paper, we show the dynamical mechanism of fusion hindrance for the $^{48}$Ca + $^{208}$Pb system by the analysis of trajectories in the three dimensional deformation space of the Langevin equation. Starting at the contact stage of colliding nuclei, we show how the delayed neck growth hinder the trajectory going toward the fusion area.  It is found that the rapid relaxation of neck makes difficult for trajectories to go up the inner slope of PES in spite of the lower barrier of the entrance stage comparing with the case of slow relaxation of neck. The point is the correlation between the evolution of inner barrier and the growth of neck during the fusion process. The dynamical variation of PES of the two center shell model and the neck formation will mainly discussed. The fission fragment mass distribution(FFMD) depending on these situation is also discussed.

In the following section, the brief review of the Langevin-type approach.
The dynamical mechanism of fusion hindrance in the $^{48}$Ca + $^{208}$Pb system at $E$$_\text{c.m.}$=180.0 MeV is shown in Sect.3, where the detail analysis on the effects of the neck formation and the reason of the fusion hindrance are presented. Our concluding remarks are given in the final section.

\section{Model}
\subsection{Potential energy surface}
We adopt the dynamical model based on the multidimensional Langevin equations, which similar to unified model \cite{Zagrebaev_2007}.
Early in the collision, the reaction stage of the nucleon transfer consists of two parts. First, at the approaching stage the system is placed in the ground state of the projectile and target because the reaction proceeds is too fast for nucleons to occupy the lowest single-particle levels. Next, the system relaxes to the ground state of the entire composite system which changes the potential energy surface (PES) to an adiabatic one.
Therefore, we treat the transition of two reaction stages with a time-dependent weighting function:
\begin{eqnarray}
&&V=V_\mathrm{{diab}}\left(q\right)f\left(t\right)+V_\mathrm{{adiab}}\left(q\right)\left[1-f\left(t\right)\right], \\
&&f\left(t\right)=\exp{\left(-\frac{t}{\tau}\right)}.
\label{pot}
\end{eqnarray}
Here, $q$ denotes a set of collective coordinates representing nuclear shape. The diabatic potential $V_{diab}\left(q\right)$ is calculated by a folding procedure using effective nucleon-nucleon interaction \cite{Zagrebaev_2005,Zagrebaev_2007,zagrebaev2007potential}.
The adiabatic potential energy $V_{adiab}\left(q\right)$ of the system is calculated using an extended two-center shell model \cite{zagrebaev2007potential}. As a characteristic of the diabatic potential, "potential wall" appears due to the overlap region of collision system which corresponds to the hard core representing the incompressibility of nuclear material.
$t$ is the interaction time and $f\left(t\right)$ is the weighting function included the relaxation time $\tau$. We use the relaxation time $\tau=10^{-22}$ s proposed in \cite{bertsch1978collision,cassing1983role,PhysRevC.69.021603}.
With the two-center parameterizations \cite{maruhn1972asymmetrie,sato1978microscopic}, the nuclear shape which represents by three deformation parameter is defined as follows:
$z_{0}$ (distance between the centers of two potentials),
$\delta$ (deformation of fragment), and $\alpha$ (mass asymmetry of colliding nuclei);
$\alpha=\frac{\left(A_{1}-A_{2}\right)}{\left(A_{1}+A_{2}\right)}$,
where $A_{1}$ and $A_{2}$ not only stand for the mass numbers of the target and projectile,
respectively \cite{Zagrebaev_2005,ARITOMO20043} but also are then used to indicate mass numbers of the two fission (heavy and light) fragments.
The parameter $\delta$ is defined as $\delta=\frac{3\left(a-b\right)}{\left(2a+b\right)}$, where $a$ and $b$ represent the half  length of the ellipse axes in the $z_{0}$ and $\rho$ directions, respectively \cite{maruhn1972asymmetrie}.
In addition, we use scaling to save computation time and use the coordinate $z$ defined as $z=\frac{z_{0}}{\left(R_{CN}B\right)}$, where $R_{CN}$ denotes the radius of the spherical compound nucleus and the parameter $B$ is defined as $B=\frac{\left(3+\delta\right)}{\left(3-2\delta\right)}$.
We solve the dynamical equation numerically. Therefore, we restricted the number of degrees of freedom as three deformation parameters to avoid the huge calculation time.

The neck parameter $\epsilon$ including in the two-center parameterizations is adjusted in Ref. \cite{YAMAJI1987487}. Reproduce the available data assuming different values between the entrance and exit channels of the reaction. In the present paper, we use $\epsilon = 1 $ for the entrance channel and $\epsilon = 0.35$ for the exit channel.
This treatment is used in Refs \cite{zagrebaev2007potential,PhysRevC.85.044614}.
We assume the time dependence of the potential energy with the finite range liquid drop model, which is denoted by the characteristic relaxation time of the neck $t_{0}$  and the variance $\Delta_{\epsilon}$ as follows:

\begin{eqnarray}
&&V_\mathrm{{LDM}}\left(q,t\right)=V_\mathrm{{LDM}}\left(q,\epsilon=1\right) f_{\epsilon}\left(t\right) \nonumber \\
&& \qquad \qquad \qquad +V_\mathrm{{LDM}}\left(q,\epsilon=0.35\right) [1-f_{\epsilon}\left(t\right)],  \\
&&V_\mathrm{{LDM}}\left(q,\epsilon\right)=E_{S}\left(q,\epsilon\right)+E_{C}\left(q,\epsilon\right), \\
&&f_{\epsilon} = \frac{1}{1+\exp\left(\frac{t-t_{0}}{\Delta_{\epsilon}}\right)},
\end{eqnarray}
where the symbols $E_{S}$ and $E_{C}$ stand for generalized surface energy and Coulomb energy, respectively \cite{PhysRevC.20.992}.
If the value of $t_{0}$ is 0 s, at the same time as contact the adiabatic potential energy for $\epsilon=1$ starts to change toward the adiabatic one for $\epsilon=0.35$. 
The time-dependent weighting function in the relaxation of $\epsilon$ value is often employed in the model based on the Langevin-type approach \cite{zagrebaev2007potential,PhysRevC.85.044614,aritomo2011dynamical,PhysRevC.96.024618,saiko2019analysis,saiko2022multinucleon}.

The adiabatic potential energy given a value of $\epsilon$ and a temperature of a system is defined as
\begin{eqnarray}
&&V_\mathrm{{adiab}}\left(q,t,L,T\right) \nonumber \\
&&\quad=V_\mathrm{{LDM}}\left(q,t\right)+V_{SH}\left(q,T\right)+V_{rot}\left(q,L\right), \\
&&V_{SH}\left(q,T\right)=E_{shell}^{0}\left(q\right)\Phi\left(T\right), \\
&&E_\text{shell}^{0}\left(q\right)=\Delta E_\text{shell}\left(q\right) + \Delta E_\text{pair}\left(q\right),\\
&&\Phi\left(T\right)=\exp\left(-\frac{E^{\ast}}{E_{d}}\right).
\label{adipot}
\end{eqnarray}
$V_{SH}$ is the shell correction energy that takes into account temperature dependence.
The symbol $E_\text{shell}^{0}$ indicates the microscopic energy at $T$ = 0, which is calculated as the sum of the shell correction energy $\Delta E_\text{shell}$ and the pairing correlation correction
energy $\Delta E_\text{pair}$. $T$ is the temperature of the compound nucleus calculated from the intrinsic energy of the composite system.
$\Delta E_\text{shell}$ is calculated by the Strutinsky method \cite{STRUTINSKY19681, RevModPhys.44.320} from the single-particle levels of the two-center shell model potential \cite{maruhn1972asymmetrie,suek74,10.1143/PTP.55.115} as the difference between the sum of single-particle energies of occupied states and the averaged quantity.
$\Delta E_\text{pair}$ is evaluated in the BCS approximation as described in Refs. \cite{RevModPhys.44.320, NILSSON19691}. The averaged part of the pairing correlation energy is calculated assuming that the density of single-particle
states is constant over the pairing window. The pairing strength constant is related to the average gap parameter $\tilde{\Delta}$ by solving the gap equation in the same approximation and adopting $\tilde{\Delta} = 12/ \sqrt{A}$ suggested in \cite{NILSSON19691} by considering the empirical results for the odd-even mass difference \cite{PhysRevC.90.054609}.
The temperature dependence factor $\Phi\left(T\right)$ is explained in Ref. \cite{ARITOMO20043}, where $E^{\ast}$ indicates the excitation energy of the compound nucleus. $E^{\ast}$ is given $E^{\ast}=aT^{2}$, where a is the level density parameter. The shell damping energy $E_{d}$ is selected as 20 MeV. This value is given by Ignatyuk et al. \cite{ignatyuk1975phenomenological}.
The rotational energy generated from the total angular momentum $L$ represents as $V_{rot}$. We obtain
\begin{eqnarray}
V_{rot}\left(q,L\right) \qquad \qquad \qquad \qquad \qquad \qquad \qquad \qquad \qquad \nonumber \\
=\frac{\hbar^{2}\ell\left(\ell+1\right)}{2\mathcal{I}\left(q\right)}+\frac{\hbar^{2}L_{1}(L_{1}+1)}{2\Im_{1}(q)}+\frac{\hbar^{2}L_{2}(L_{2}+1)}{2\Im_{2}(q)}. \quad
\end{eqnarray}
Here, $\mathcal{I}\left(q\right)$ and $\ell$ represent the moment of inertia of the rigid body with deformation $q$ and the relative orientation of nuclei and relative angular momentum respectively. The moment of inertia and the angular momentum  for the heavy and light fragments are $\Im_{1,2}$ and $L_{1,2}$, respectively.


\subsection{Dynamical equations}
The trajectory calculations are performed on the time-dependent unified potential energy \cite{Zagrebaev_2005,Zagrebaev_2007,ARITOMO20043} using the multidimensional Langevin equation \cite{Zagrebaev_2005,ARITOMO20043,PhysRevC.80.064604} as follows:
\begin{gather}
\frac{dq_{i}}{dt}=\left(m^{-1}\right)_{ij}p_{j}, \nonumber \\
\frac{dp_{i}}{dt}=-\frac{\partial V}{\partial q_{i}}-\frac{1}{2}\frac{\partial}{\partial q_{i}}\left(m^{-1}\right)_{jk}p_{j}p_{k}-\gamma_{ij}\left(m^{-1}\right)_{jk}p_{k} \nonumber \\
+g_{ij}R_{j}\left(t\right), \nonumber \\
\frac{d\theta}{dt}=\frac{\ell}{\mu_{R}R^{2}}, \nonumber \quad
\frac{d\varphi_{1}}{dt}=\frac{L_{1}}{\Im_{1}}, \nonumber \quad
\frac{d\varphi_{2}}{dt}=\frac{L_{2}}{\Im_{2}}, \nonumber \\
\frac{d\ell}{dt}=-\frac{\partial V}{\partial\theta}-\gamma_{tan}\left(\frac{\ell}{\mu_{R}R^{2}}-\frac{L_{1}}{\Im_{1}}a_{1}-\frac{L_{2}}{\Im_{2}}a_{2}\right)R
\nonumber \\
+Rg_{tan}R_{tan}\left(t\right), \nonumber \\
\frac{dL_{1}}{dt}=-\frac{\partial V}{\partial\varphi_{1}}+\gamma_{tan}\left(\frac{\ell}{\mu_{R}R^{2}}-\frac{L_{1}}{\Im_{1}}a_{1}-\frac{L_{2}}{\Im_{2}}a_{2}\right)a_{1}  \nonumber \\
-a_{1}g_{tan}R_{tan}\left(t\right), \nonumber \\
\frac{dL_{2}}{dt}=-\frac{\partial V}{\partial\varphi_{2}}+\gamma_{tan}\left(\frac{\ell}{\mu_{R}R^{2}}-\frac{L_{1}}{\Im_{1}}a_{1}-\frac{L_{2}}{\Im_{2}}a_{2}\right)a_{2}
\nonumber \\
-a_{2}g_{tan}R_{tan}\left(t\right).
\end{gather}
The collective coordinates $q_{i}$ represent $z, \delta$, and $\alpha,$ the symbol $p_{i}$ denotes momentum conjugated to $q_{i}$, and $V$ is the multidimensional potential energy. The symbol $\theta$ indicates the relative orientation of nuclei. $\varphi_{1}$ and $\varphi_{2}$ stand for the rotation angles of the nuclei in the reaction plane, $a_{1,2}=\frac{R}{2}\pm\frac{R_{1}-R_{2}}{2}$ is the distance from the center of the fragment to the middle point between the nuclear surfaces, and $R_{1,2}$ is the nuclear radii. The symbol $R$ is the distance between the nuclear centers.
The total angular momentum $L=\ell+L_{1}+L_{2}$ is preserved. The symbol $\mu_{R}$ is reduced mass, and $\gamma_\text{tan}$ is the tangential friction force of the colliding nuclei.
The phenomenological nuclear friction forces for separated nuclei are expressed in terms of the tangential friction $\gamma_\text{tan}$ and the radial friction $\gamma_{R}$ using the Woods-Saxon radial form factor described in Ref. \cite{Zagrebaev_2005} as follows:
\begin{eqnarray}
&&F\left(\xi\right)=\left(1+\exp^{\frac{\xi-\rho_{F}}{a_{F}}}\right)^{-1},\\
&&\gamma_\text{tan}=\gamma_\text{t}^{0}F\left(\xi\right),\\
&&\gamma_{R}=\gamma_{R}^{0}F\left(\xi\right).
\end{eqnarray}
The model parameter $\gamma_\text{t}^{0}$ and $\gamma_{R}^{0}$ which is used in the previous paper \cite{PhysRevC.106.024610} employ 0.1 $\times~10^{-22}$ MeV s fm$^{-2}$ and 100 $\times~10^{-22}$ MeV s fm$^{-2}$, respectively. $\rho_{F}$ and $a_{F}$ are 2 fm and 0.6 fm which are determined in Ref. \cite{Zagrebaev_2005}. $\xi$ is the distance between the nuclear surfaces $\xi=R-R_\text{contact}$, where $R_\text{contact}=R_{1}+R_{2}$.
The phenomenological friction for the radial direction is switched to the one-body friction in the mononucleus state. $\gamma_{R}$ is described to consider the kinetic dissipation according to the surface friction model \cite{FROBRICH1998131}. The radial friction is calculated as
\begin{eqnarray}
\gamma_{zz}=\gamma_{zz}^\text{{one}}+\Omega\left(\xi\right)\gamma_{R}.
\end{eqnarray}
For the mononuclear system, the wall-and-window one-body dissipation $\gamma_{zz}^\text{{one}}$ is adopted for the friction tensor \cite{BLOCKI1978330,RAYFORDNIX1984161,RANDRUP1984105,Feldmeier_1987,CARJAN1986381,carj92,PhysRevLett.70.3538,20067}.
The phenomenological friction is switched to that of a mononuclear system using the smoothing function \cite{Zagrebaev_2005}
 \begin{eqnarray}
\Omega\left(\xi\right)=\left(1+\exp^{-\frac{\xi}{0.3}}\right)^{-1}.
\end{eqnarray}
$m_{ij}$ and $\gamma_{ij}$ stand for the shape-dependent collective inertia and friction tensors, respectively.
We adopted the hydrodynamical inertia tensor $m_{ij}$ in the Werner-Wheeler approximation for the velocity field \cite{PhysRevC.13.2385}.
The one-body friction tensors $\gamma_{ij}$ are evaluated within the wall-and-window formula \cite{RANDRUP1984105, PhysRevC.21.982}.
The normalized random force $R_{i}\left(t\right)$ is assumed to be white noise: $\langle R_{i} (t) \rangle$ = 0 and $\langle R_{i} (t_{1})R_{j} (t_{2})\rangle = 2 \delta_{ij}\delta (t_{1}-t_{2})$.
According to the Einstein relation, the strength of the random force $g_{ij}$ is given as $\gamma_{ij}T=\sum_{k}{g_{ij}g_{jk}}$.


\section{Results}
\subsection{Fusion hindrance due to neck formation}

\begin{figure}[t]
\centering
\includegraphics[scale=0.34]{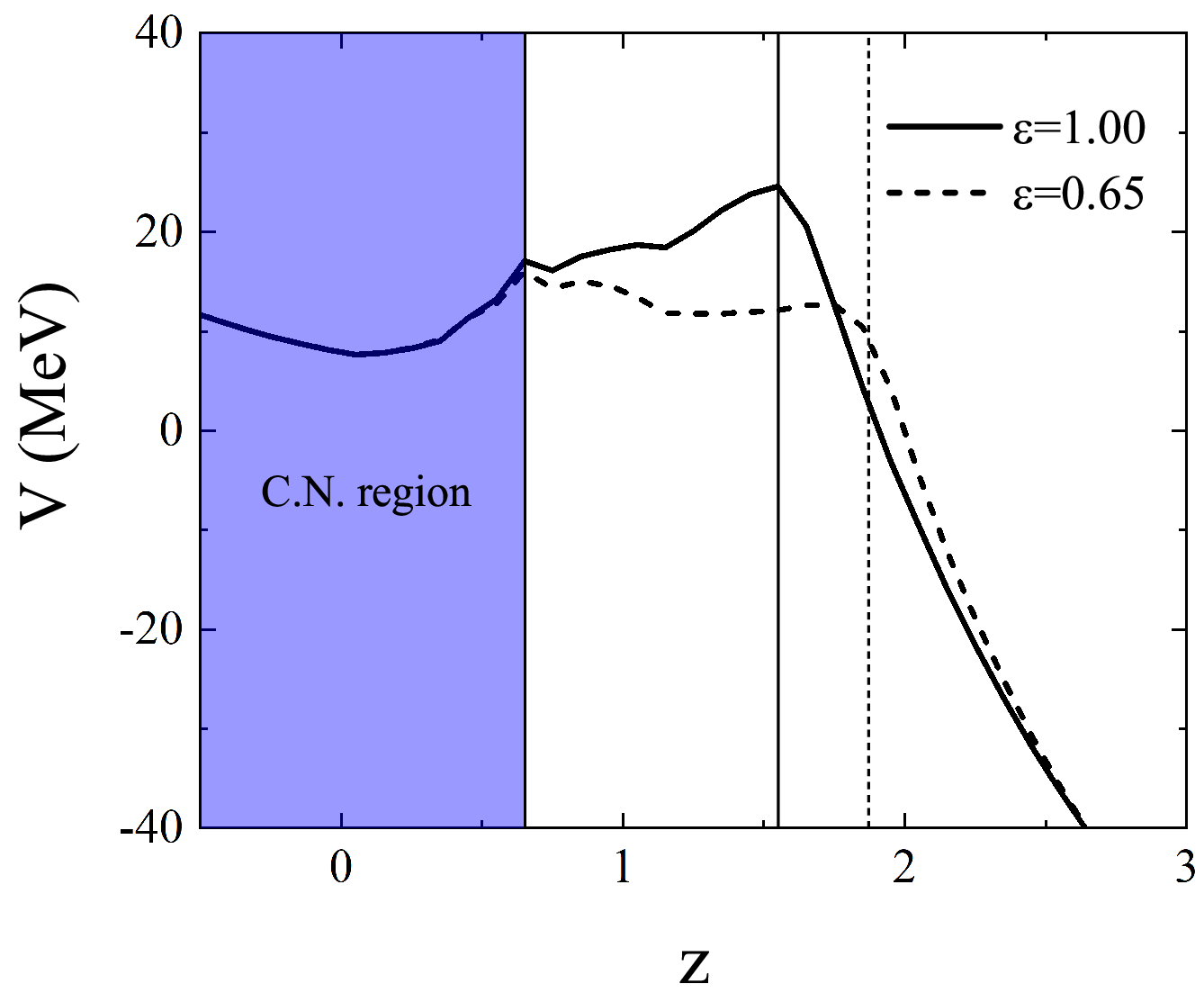}
\caption{One-dimensional fusion barrier for the central collision ($L=0\hbar$) in the reaction $^{48}$Ca+$^{208}$Pb. The solid and dashed lines are for fixed $\epsilon=1.00$ and fixed $\epsilon=0.65$, respectively. The $z$ value at the contact without neck bridge formation and with neck bridge formation are indicated by vertical solid and dashed lines. The blue region shows the compound nucleus (CN) region.
}\label{fig1}
\end{figure}

\begin{figure}[t]
\centering
\includegraphics[scale=0.34]{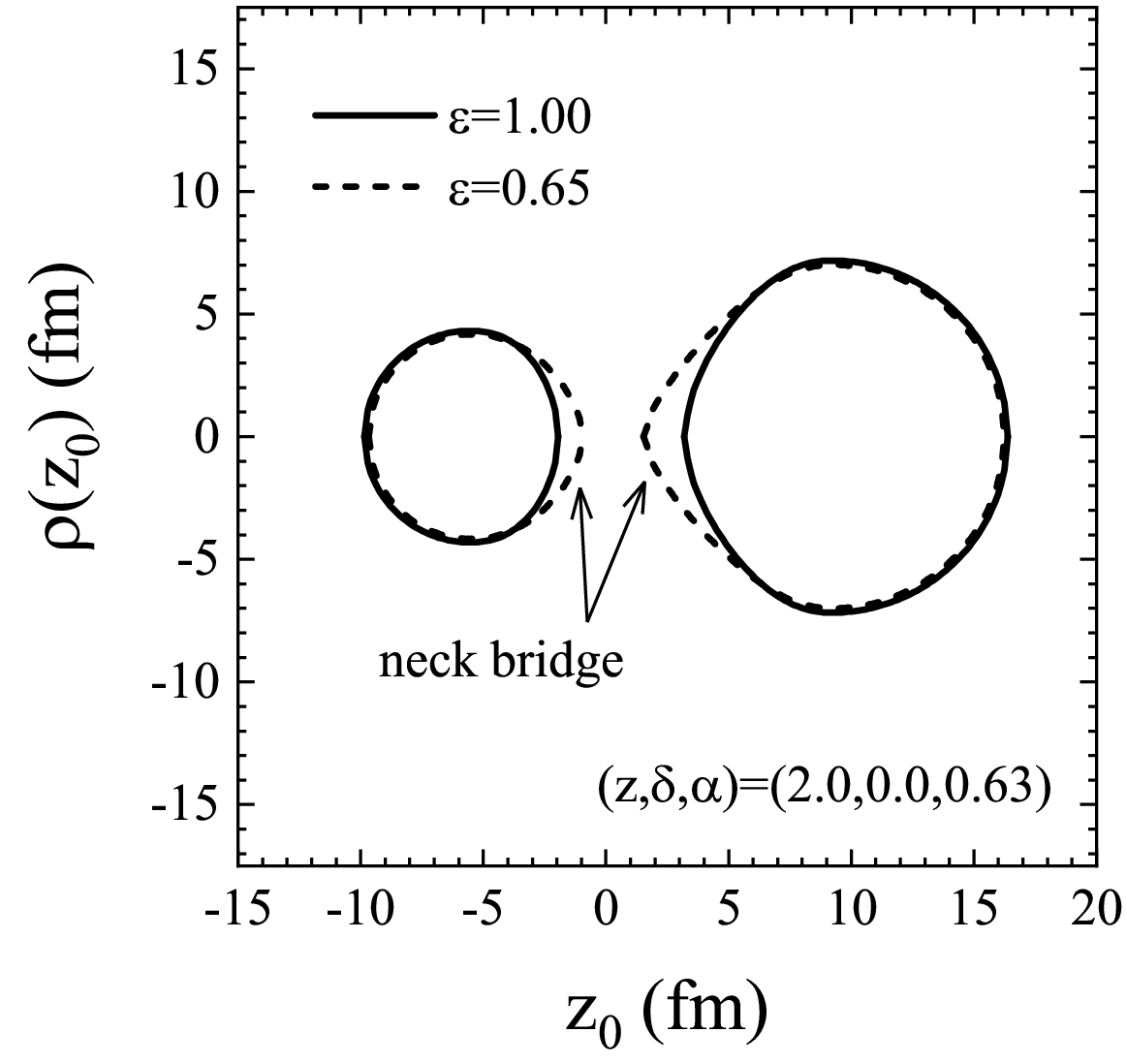}
\caption{Nuclear configuration for $^{48}$Ca+$^{208}$Pb before two nuclei contact (\{$z$,$\delta$,$\alpha$\} = \{2.0,0.00,0.63\}). The solid and dashed lines correspond to nuclear configurations for $\epsilon=1.00$ and $\epsilon=0.65$, respectively.
}\label{fig2}
\end{figure}

\begin{figure}[t]
\centering
\includegraphics[scale=0.35]{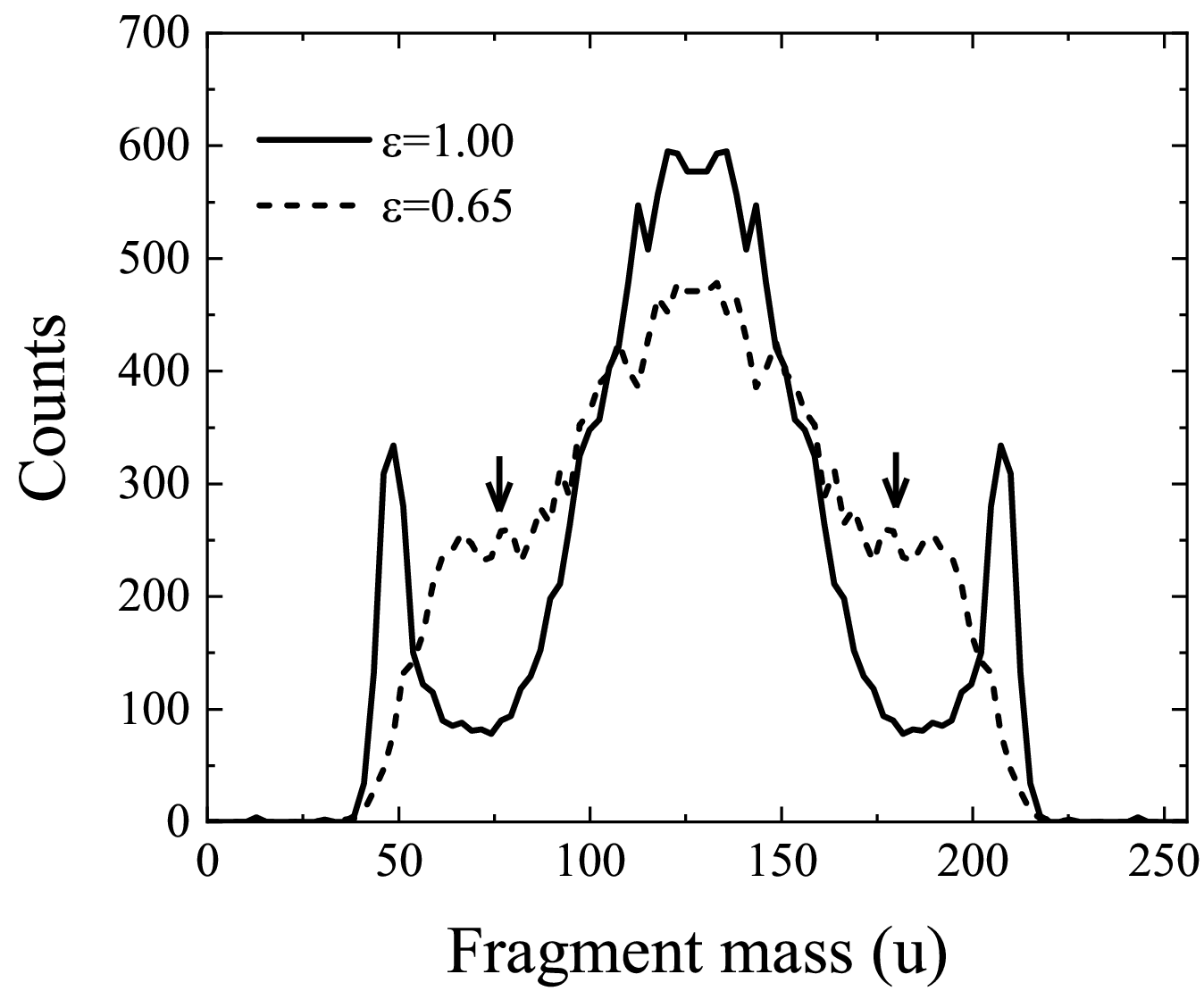}
\caption{Mass distribution of fission fragments for the central collision ($L=0\hbar$) in the reaction $^{48}$Ca+$^{208}$Pb at $E$$_\text{c.m.}$ = 180.0~MeV. The solid and dashed lines show the calculation results for fixed $\epsilon=1.00$ and fixed $\epsilon=0.65$, respectively.
}\label{fig3}
\end{figure}

\begin{figure*}[t]
\centering
\includegraphics[scale=0.65]{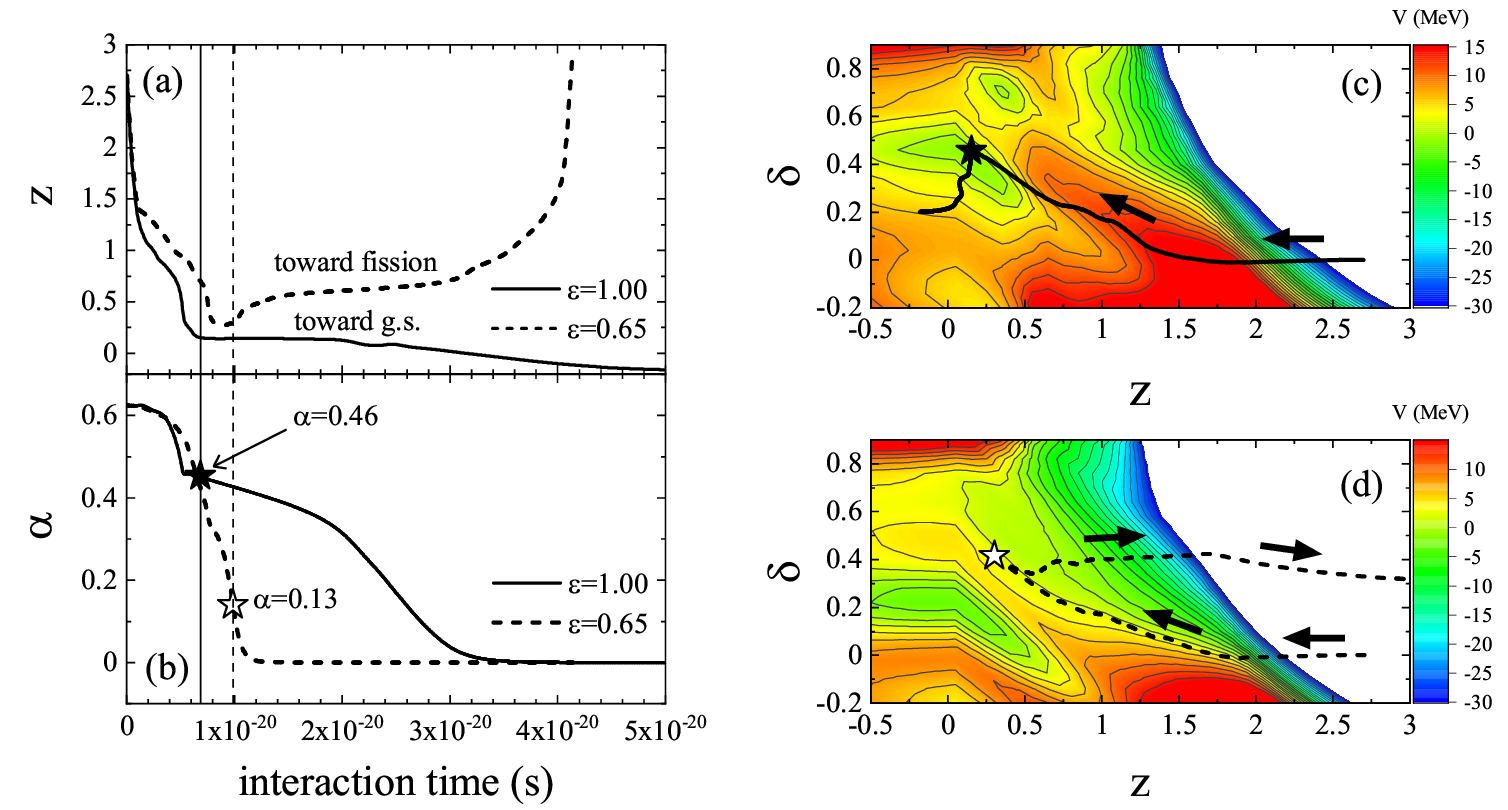}
\caption{The temporal evolution without fluctuation of $z$ (a) and $\alpha$ (b) for the different $\epsilon$ value. The significant time is indicated by vertical solid and dashed lines. The solid and open star points show the significant value of mass asymmetry for fixed $\epsilon=1.00$ and fixed $\epsilon=0.65$, respectively.
The potential energy surface of $z$-$\delta$ plane at the solid star point ($\alpha=0.46$) (c) and the open star point ($\alpha=0.13$) (d).
The mean trajectory drawn in potential energy surface of $z-\delta$ plane at the solid star point ($\alpha=0.46$) for fixed $\epsilon=1.00$ (c). The same as (c), but at the open star point ($\alpha=0.13$) for fixed $\epsilon=0.65$ (d).
}\label{fig4}
\end{figure*}

%
To review the fusion hindrance due to effects of the neck formation in heavy ion collisions, the one-dimensional fusion barriers for the fixed $\epsilon$ (1.0 and 0.65)  are shown for the reaction of $^{48}$Ca+$^{208}$Pb in Fig. 1. The top of barrier and the contact line are at the same position in the case of $\epsilon=1.0$. In other words, the inner barrier is not formed for $\epsilon=1.0$. Thus, the system overcoming the fusion barrier after the contact of two nuclei moves easily toward the formation of a compound nucleus.
On the other hand, when two nuclei form a neck bridge before contact, the fusion barrier decreases at the contact point. The fusion enhancement may be expected due to the decrease of barrier height, however the system needs to overcome the inner barrier against the friction forces and as a result the fusion hindrance occurs. Details will be discussed later.

%
The nuclear configuration before contact point \{$z,\delta,\alpha$\}=\{2.0,0.0,0.63\} for different fixed $\epsilon$ value in $^{48}$Ca+$^{208}$Pb is shown in Fig. 2. As shown in the figure, if use is made of $\epsilon=0.65$, the neck bridge is formed before contact.

%
Next, we investigate the fission fragment mass distribution (FFMD) under the different neck bridge formation in the reaction $^{48}$Ca+$^{208}$Pb at $E$$_\text{c.m.}$ = 180.0~MeV. FFMD without the effect of the orbital angular momentum for $\epsilon=1.0$ and $\epsilon=0.65$ is shown in Fig. 3. 
The mass symmetric fission decreases for the case of the neck bridge formation ($\epsilon=0.65$) before contact comparing with that for the case of $\epsilon=1.0$. The decrease of the mass symmetric fission is due to the inner barrier after contact. The sharp peaks at both side of the distribution are the events of the quasi-elastic (QE) reaction, which are confirmed in the mass of the collision system. The formation of the neck bridge before contact makes a second bump indicated by arrows near $^{180}$Hg due to the shell effects. These bumps correspond to the quasi-fission (QF) events keeping the  memory of entrance channel.

%
We also investigate the FFMD when the formation of the neck bridge appears before contact in terms of the dynamical analysis using the mean trajectory calculations. Figure 4 shows the calculation result of the $^{48}$Ca+$^{208}$Pb reaction without the effect of the angular momentum at $E$$_\text{c.m.}$ = 180.0 MeV. The calculation starts at \{$z,\delta,\alpha$\}=\{2.65,0.0,0.63\}. 
Both of the temporal evolution of $z$ with and without the formation of the neck bridge before contact are shown in Fig. 4(a). The value of $z$ begins to move toward the ground state for no neck bridge formation as indicated by the solid line. If the neck bridge is formed before contact, $z$ goes toward the fission area as shown by the broken line. The characteristic times at which the dynamical variation for $\epsilon=1.0$ and $\epsilon=0.65$ are shown by vertical solid and dashed lines, respectively. Each time is 6.845$\times10^{-21}$ s and 9.873$\times10^{-21}$ s.

Figure 4(b) shows the calculation result of mean temporal evolution of $\alpha$. When the neck bridge is formed before contact ($\epsilon=0.65$), the system can obtain the sufficient neck cross section. The relaxation of the degree of mass asymmetry starts rapidly in an early stage by nucleon transfer through the neck. On the contrary in the case of $\epsilon=1.0$, the relaxation of degree of mass asymmetry delays because the thick neck is not formed. Note that the fluctuations are not taken account in these calculation. 
The characteristic location where the dynamic changes occur are indicated by black star ($\alpha=0.46$) and white star ($\alpha=0.13$) in Fig.~4(b).

We analyze the dynamical difference of these two temporal evolution in $\alpha$. The mean trajectories for $\epsilon=1.0$ and $\epsilon=0.65$ drawn in $z-\delta$ plane of potential energy surface at the point of $\alpha=0.46$ and $\alpha=0.13$ are shown in Fig. 4(c) and Fig. 4(d), respectively. The black and white stars plotted in Fig. 4(c) and Fig. 4(d) correspond to the locations of the black and white stars in Fig. 4(b), respectively. From the behavior of the trajectory passing after the black star in Fig. 4(c), the trajectory is guided to the ground state because no barrier exists toward the fusion area.
On the other hand, the trajectory returns back at the white star in Fig. 4(d) to the fission area, because of the inner barrier appeared due to the rapid relaxation of the degree of mass asymmetry. Therefore, the trajectory cannot invade the CN region and then goes in the direction of the fission. The formation of the inner barrier is caused by the rapid relaxation of the degree of mass asymmetry due to the formation of a sufficiently thick neck at the early time (case $\epsilon=0.65$). It can be seen that the quasi-fission is controlled by the initial dynamics of the neck.

%
%
%
%
%
\subsection{FFMD depending on the neck relaxation}\label{sec3B}

\begin{figure}[t]
\centering
\includegraphics[scale=0.3]{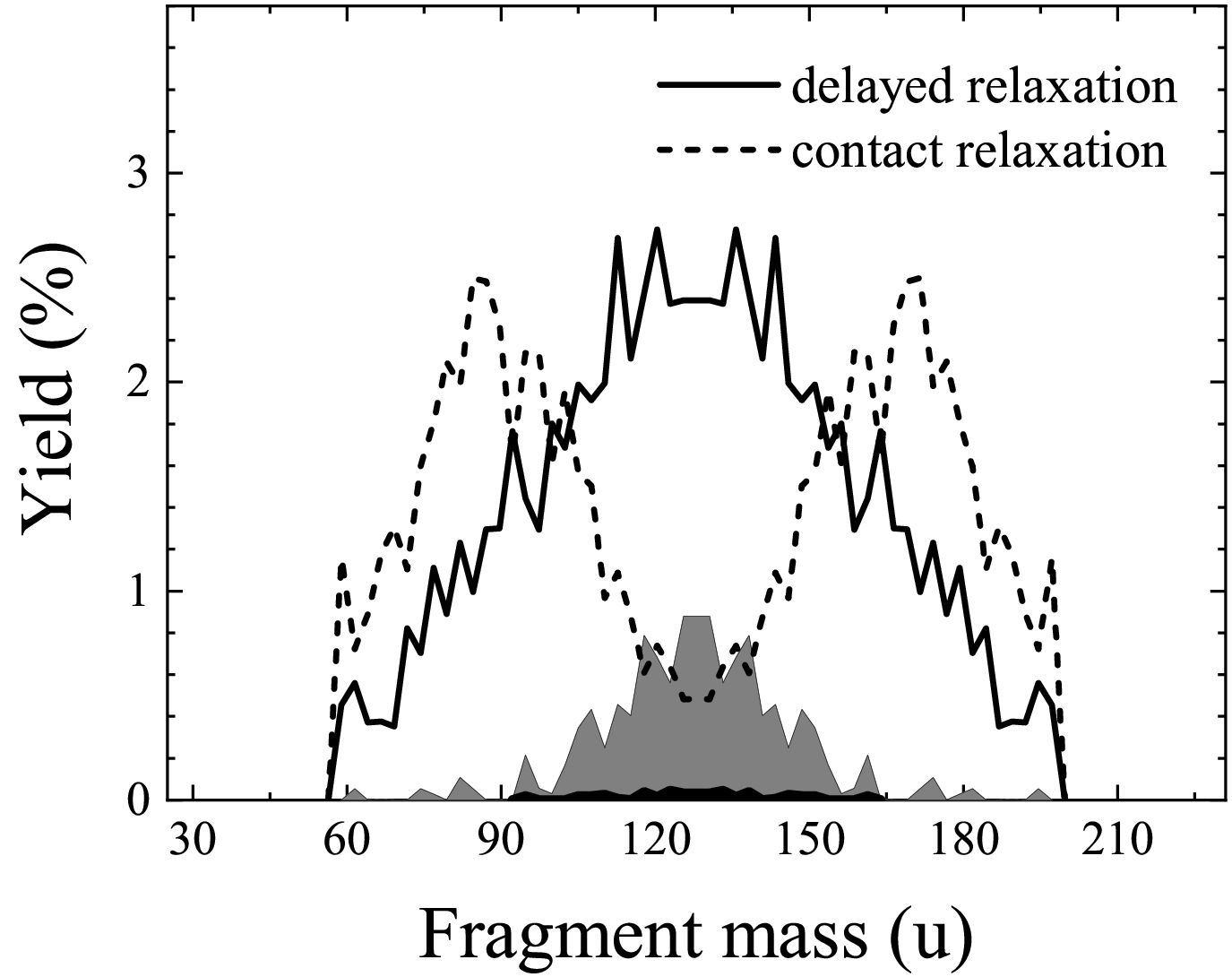}
\caption{Mass distribution of fission fragments for the initial orbital angular momentum in the range of $L=0-100$~$\hbar$ for the reaction $^{48}$Ca+$^{208}$Pb at $E$$_\text{c.m.}$ = 180.0~MeV. The solid and dashed lines show the calculation results of two relaxation mode of $\epsilon$: the relaxation starting late after contact and immediately after contact, respectively. The fusion-fission fragments for ``delayed relaxation'' and ``contact relaxation'' are presented by the grey and the black-filled areas, respectively. 
}\label{fig5}
\end{figure}

\begin{figure*}[t]
\centering
\includegraphics[scale=0.6]{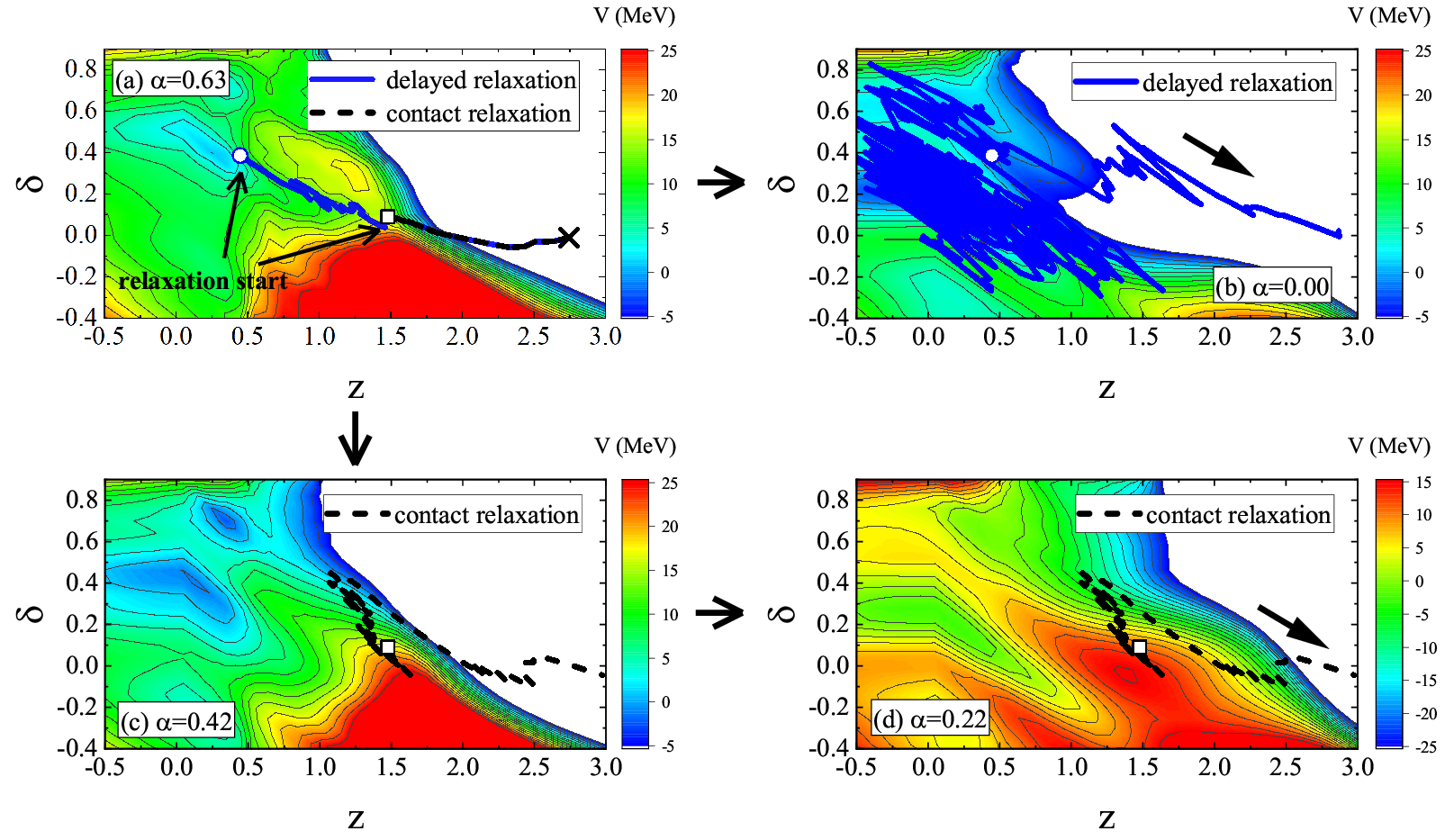}
\caption{The trajectories with fluctuation drawn in potential energy surface for each $\alpha$ value in the reaction $^{48}$Ca+$^{208}$Pb at $E$$_\text{c.m.}$ = 180.0~MeV. The solid blue and dashed black lines indicate $\epsilon$ relaxation starting late after contact and immediately after contact, respectively. The $\times$ are calculation starting point. The relaxation starting points of $\epsilon$ parameter are indicated by black square and blue circle. The trajectories up to the onset of relaxation are drawn in (a) and the trajectories after the onset of relaxation are drawn in (b)-(d). Black arrows indicate the direction of the procession of trajectories toward fission.
}\label{fig6}
\end{figure*}

\begin{figure}[t]
\centering
\includegraphics[scale=0.34]{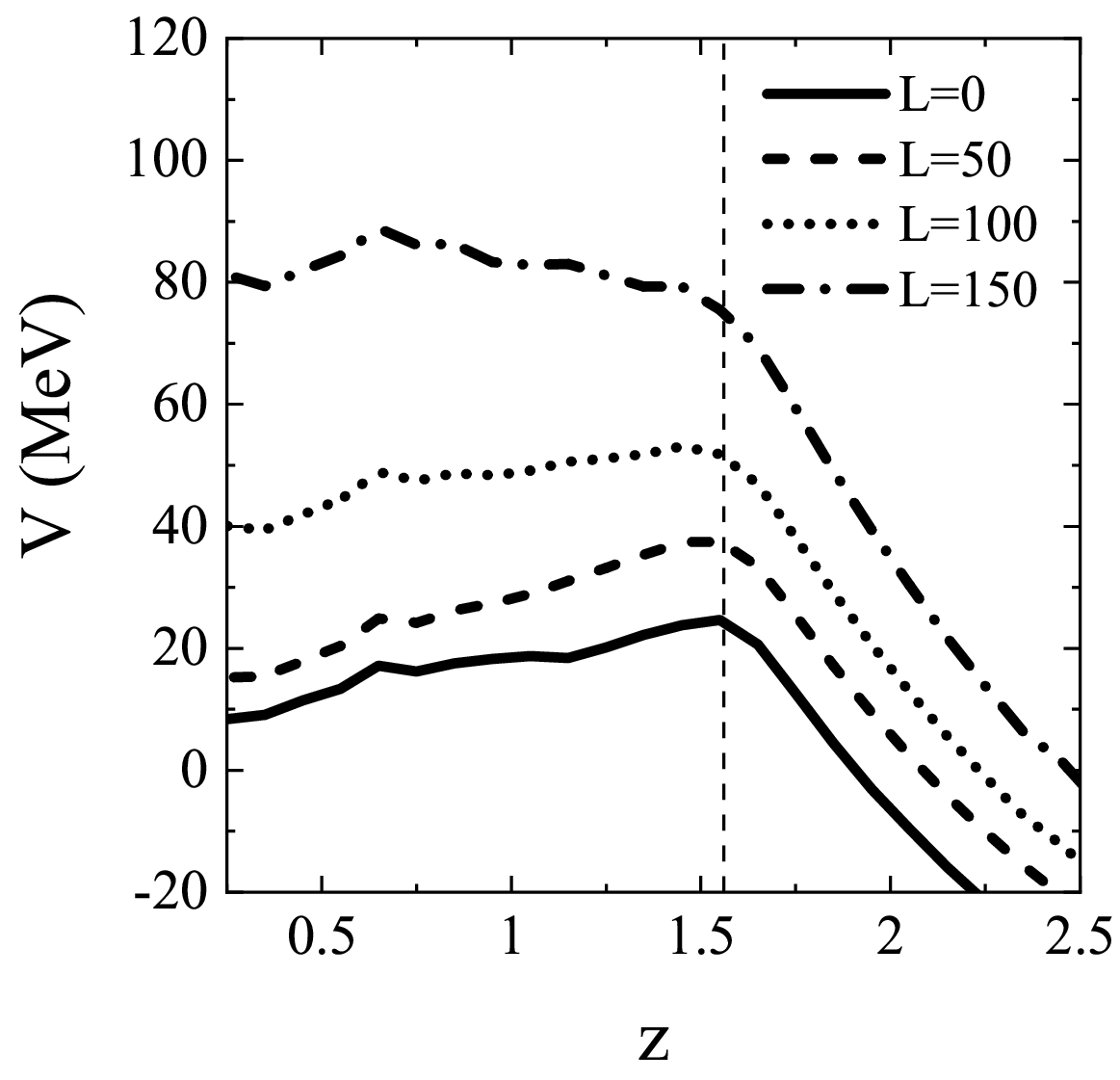}
\caption{One-dimensional fusion barrier depending on the initial angular momentum ($L$) for a fixed value of $\epsilon=1.0$ in the reaction $^{48}$Ca+$^{208}$Pb. The solid, dashed, dotted, and dotted-dashed lines are for $L=0$, 50, 100, and 150 $\hbar$, respectively. The $z$ value at the contact is indicated by vertical dashed lines.
}\label{fig7}
\end{figure}

\begin{figure*}[t]
\centering
\includegraphics[scale=0.6]{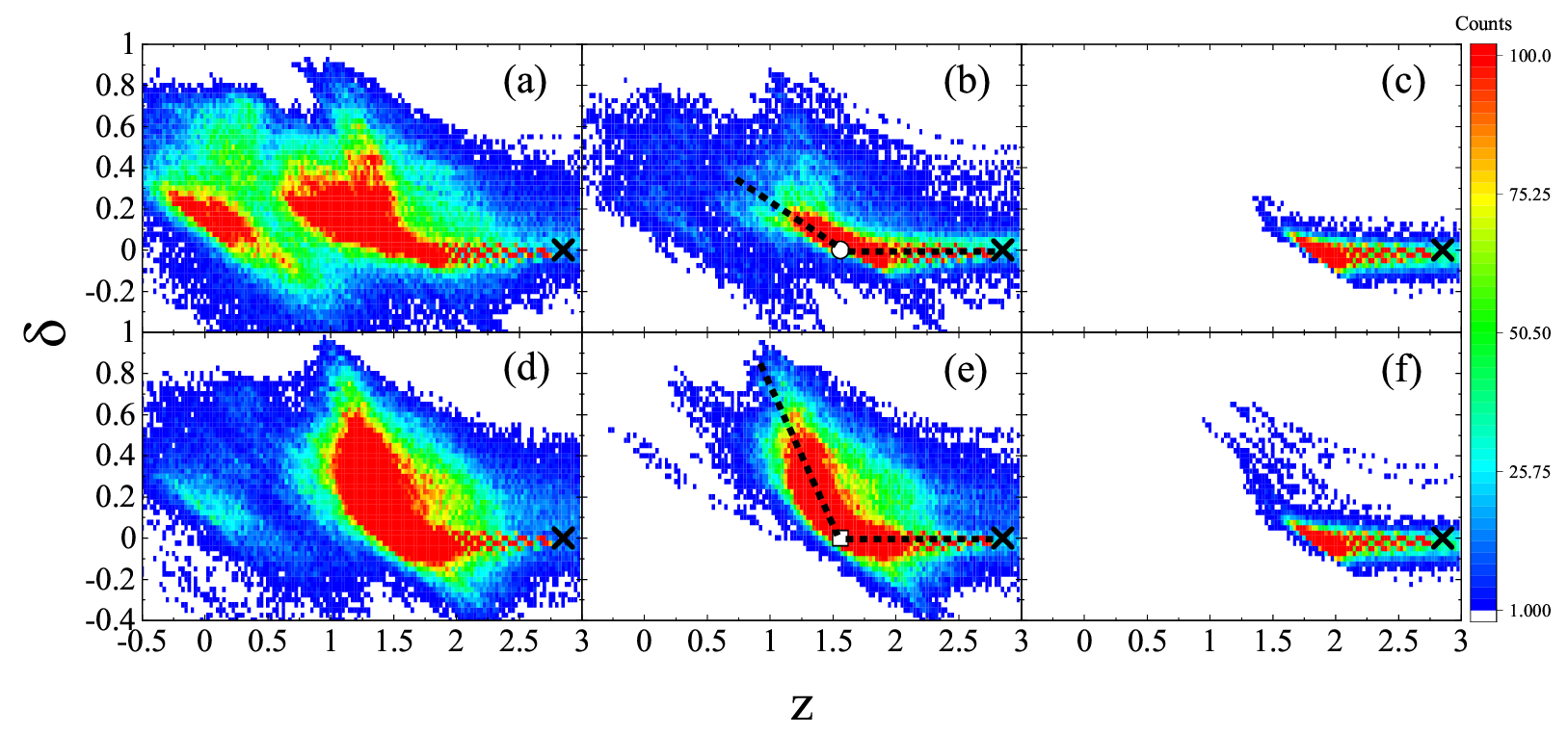}
\caption{Trajectory distribution at the different initial angular momentum and relaxation modes drawn in $z-\delta$ plane for $^{48}$Ca+$^{208}$Pb system with $E$$_\text{c.m.}$ = 180.0 MeV. The $\times$ are the starting point of calculation. Panels (a)-(c) and (d)-(f) show the trajectory distribution for ``delayed relaxation'' and ``contact relaxation'' of $\epsilon$, respectively. Panels (a)(d), (b)(e) and (c)(f) represent the results in the initial angular momentum 0$\hbar$, 30$\hbar$ and 60$\hbar$, respectively. The open circle and square show the contact point in (b) and (e), respectively. The dashed lines are drawn to guide the direction of trajectories from the starting point.
}\label{fig8}
\end{figure*}

\begin{figure}[t]
\centering
\includegraphics[scale=0.34]{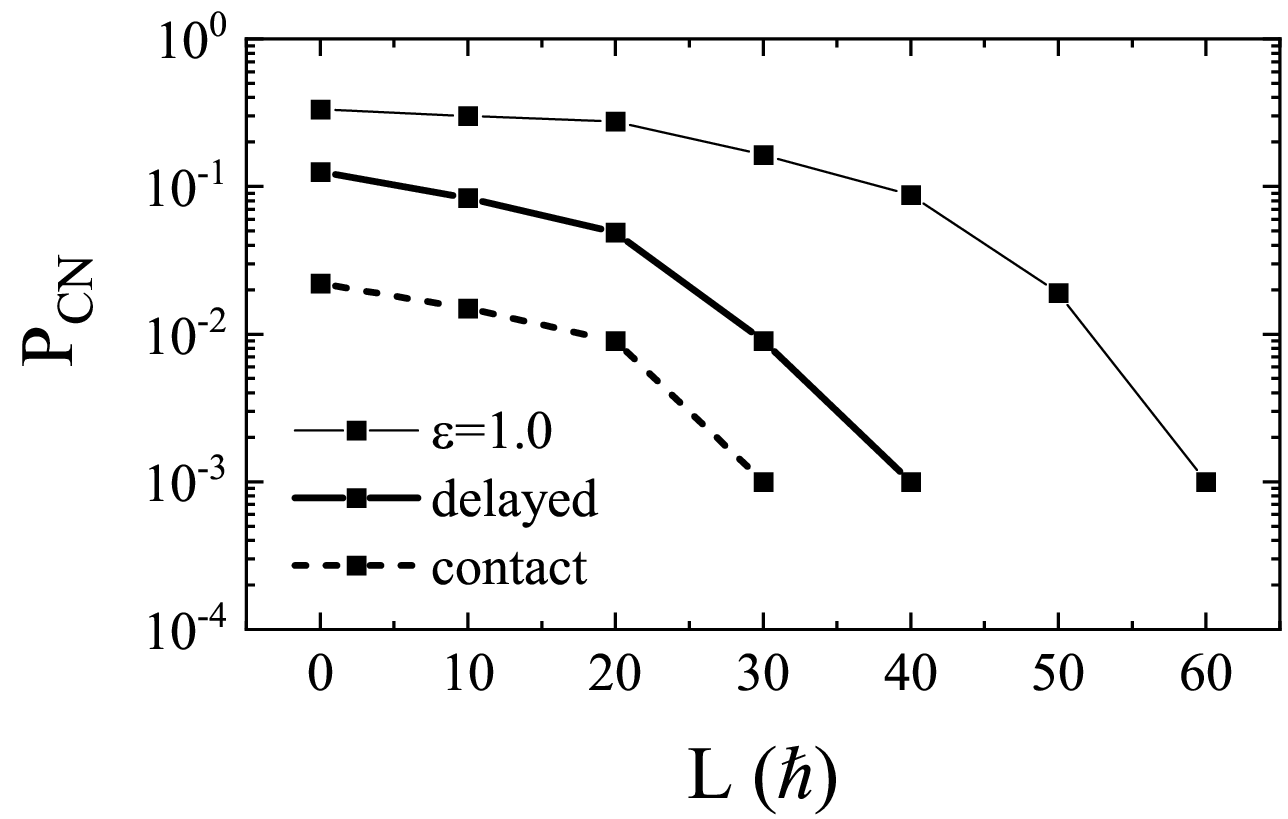}
\caption{Fusion probability for $^{48}$Ca+$^{208}$Pb at $E$$_\text{c.m.}$ = 180.0 MeV to various $L$ values. The thick solid line is the case of $\epsilon$ relaxation starting late after contact and the dashed line is for $\epsilon$ relaxation starting immediately at the two nuclei contact. The thin line is the calculations for a fixed value of $\epsilon=1.0$.
}\label{fig8}
\end{figure}

%
We investigate the dynamics of $\epsilon$ in the Langevin calculation in connection with FFMD. We assume that the relaxation of $\epsilon$ is generated by using $\Delta_{\epsilon}=1.0\times10^{-22}$ s in eq.(5). Theoretical reports \cite{ADAMIAN199929,ADAMIAN2000233,zhao09,abe2010,boilley2011} show that once $\epsilon$ starts to relax, the relaxation proceeds rapidly. The starting time of the relaxation is adjusted by $t_{0}$. To investigate the different initial dynamics of $\epsilon$, we use $t_{0}=0$ s and $5.0\times10^{-21}$ s. We call each relaxation mode as ``contact relaxation'' and ``delayed relaxation''.
Figure 5 shows the FFMD calculated for the different relaxation modes in $^{48}$Ca+$^{208}$Pb reaction at $E$$_\text{c.m.}$ = 180.0MeV. We cut the events in the mass range $A<58$, $198<A$ where quasi-elastic collision is dominated. The mass asymmetric fission is dominant for ``contact relaxation'', and the mass symmetric fission is dominant for ``delayed relaxation''. 
The inner barrier is formed by the early formation of the neck in the case of ``contact relaxation'', however, the fusion yield is slightly ensured because trajectories overcome the inner barrier due to the effect of dynamical fluctuation. 
The percentage of fusion-fission is 0.57 \%(black filled part) and 14.23 \%(grey filled part) for ``contact relaxation'' and ``delayed relaxation'', respectively. It is clear the CN formation cross section is hindered for ``contact relaxation'' mode. We analyze the difference of the dominant fission mode dynamically.

%
 The sample trajectory of  ``contact relaxation'' and ``delayed relaxation'' with a fluctuation are drawn in the $z-\delta$ potential energy surface at each $\alpha$ in Fig. 6. The calculation starts at the $\times$. As can be seen in Fig.~6(a), the trajectories go up to the black square and the blue circle where the relaxation of $\epsilon$ starts in. The trajectories of  ``contact relaxation'' and ``delayed relaxation'' keep $\epsilon$ to 1.0 from the $\times$ to the black square and the blue circle, respectively.

In the case of ``delayed relaxation'', the trajectory follows from the black square in Fig.~6(a) to the blue circle without neck relaxation. The blue circle is also indicated in Fig.~6(b), but the PES changes drastically because of the relaxation of mass asymmetry described below. 
The trajectory moves toward the compound nucleus region due to the inner slope like in Fig.~1 (solid line) corresponding to the case of ``delayed relaxation''. After that, since the trajectory is trapped in the pocket which appeared around \{$z,\delta$\}=\{0.0,0.2\}, the mass asymmetry is rapidly relaxed. The effect of this pocket appearing in the deformation space has been already reported in Ref. \cite{ARITOMO2005152}. Then the trajectory with sufficiently relaxed in $\alpha$ exits from near $\alpha=0$ as indicated by the black arrow. 
In Fig~4(c), the trajectory settles down the pocket which appeared around \{$z,\delta$\}=\{0.0,0.2\}, however, the trajectory does not settle down the pocket in Fig~6(b). Finally, trajectory moves to fission due to the effect of dynamical fluctuation. 
In the case of ``contact relaxation'', the mass transfer through the neck occurs actively before overcoming the inner barrier.
The trajectory and PES change from Fig.~6(a) to Fig.~6(c) due to the early relaxation of $\alpha$. As can be seen in Fig.~6(c), it is difficult for the trajectory to enter the compound nucleus region by the formation of the inner barrier due to early relaxation of the mass asymmetry. Finally, the path to fusion is hindered and the system goes toward fission at $\alpha=0.2$ (as the black arrow in Fig.~6(d)). The same trajectory is drawn in Fig.~6(c) and Fig.~6(d).

\subsection{Denecking process with the initial orbital angular momentum}

%
Figure 7 shows the one-dimensional fusion barrier depending on the initial angular momentum in $^{48}$Ca+$^{208}$Pb reaction. The barrier at the contact point becomes higher according to the centrifugal potential energy. Therefore, the formation of CN is expected to suppress corresponding to $L$.

%
We investigate how the discussion presented above for $L=0$~$\hbar$ is modified in the case with the angular momentum in the reaction of $^{48}$Ca+$^{208}$Pb at $E$$_\text{c.m.}=180.0~\text{MeV}$. The trajectory distributions for several $L$ value on $z-\delta$ plane are shown in Fig.~8. The upper three panels Figs.~8(a)-(c) are for ``delayed relaxation'' of $\epsilon$ and the lower three Figs.~8(d)-(f) are for ``contact relaxation'' of $\epsilon$. All trajectories start at $\times$. In the case of ``delayed relaxation'', substantial trajectories for $L=0$~$\hbar$ and $30$~$\hbar$ reach to the compound nucleus region \{$z,\delta$\}=\{0.0,0.2\}, but in the case of ``contact relaxation'', trajectories coming to the compound nucleus region are limited only to $L=0$~$\hbar$. As shown in Fig.~6(a),  the distribution is enhanced at the pocket of \{$z,\delta$\}=\{0.0,0.2\}, where the relaxation of $\alpha$ occurs rapidly, and the mass symmetric fission becomes dominant. The trajectories for $L=60$~$\hbar$ cannot overcome the barrier due to the centrifugal potential.

The mean direction of the motion of trajectory after contact for ``delayed relaxation''  is clearly different from that for ``contact relaxation''. As can be seen in Fig.~8(b), the inclination of the trajectories has about $30^\circ$ from the horizon line after contact. However, the inclination of the trajectories for ``contact relaxation'' has $60^\circ$ as shown in Fig.~8(e). This trend also occurs for other orbital angular momentum. This difference comes from forming the inner barrier due to the early growth of the neck described in \ref{sec3B}, and is the main factor for the fusion hindrance.

%
Finally, we try to estimate the fusion probability ($P_\text{CN}$). The definition of fusion probability in our model is given in Refs. \cite{ARITOMO20043,PhysRevC.85.044614}. Figure 9 shows $P_\text{CN}$ of each orbital angular momentum for the different relaxation modes of $\epsilon$ in $^{48}$Ca+$^{208}$Pb reaction at $E$$_\text{c.m.}$ = 180.0 MeV. $P_\text{CN}$ is the highest when $\epsilon$ is fixed to 1.0. This is due to no formation of inner barrier by the relaxation of $\epsilon$. $P_\text{CN}$ for ``delayed relaxation'' is higher by one order than that for ``contact relaxation'', and extends to $L=40$~$\hbar$. It is clear that the fusion enhancement is expected for ``delayed relaxation'' of $\epsilon$. The early relaxation of $\epsilon$ occurs the fusion hindrance by the formation of inner barrier. The time when the denecking motion (relaxation of $\epsilon$) starts is the critical matter for the estimation of fusion probability in superheavy nuclei.

\section{Summary}
The fusion hindrance due to the formation of the neck has been investigated using the dynamical model based on Langevin equations. The fusion barrier decreases by forming the neck bridge at the approaching stage. However, if the neck bridge is formed before contact, the events of the mass symmetric fission accumulated inside the fragment mass of $A$$_\text{CN}$$/2 \pm$20u are decreased.
The fusion probability is suppressed for the early rapid relaxation of $\epsilon$ than the case for the delayed relaxation of $\epsilon$. As the results of the trajectory analysis on the nuclear deformation space, it is found that if the neck relaxation starts in the early stage of collision, the fusion barrier decreases as a whole but the uphill inner barrier arises, and the trajectory is prevented from going inside the fusion area. Therefore, it is concluded that the fusion hindrance comes from the formation of the inner barrier due to the early denecking process.
In heavy ion collisions, the diabatic PES is adopted from the approaching process to the initial stage of contact.
In the early stage of collision, the time dependent function of the neck relaxation should be treated precisely by considering the diabatic situation of the system.
 
%

\begin{acknowledgments}
The Langevin calculation were performed using the cluster computer system (Kindai-VOSTOK) which is supported by JSPS KAKENHI Grant Number 20K04003 and Research funds for External Fund Introduction 2021 by Kindai University.
\end{acknowledgments}
%
%


\end{document}